%% file: OneSided.tex
\pdfoutput=1
\documentclass[usegraphicx,usenatbib]{mn2e}

\usepackage{amssymb,bm,color,charter}
\usepackage{ulem}
\input{shortcuts.tex}

\title[One-sided Jets]
{One-sided Outflows/Jets from Rotating Stars with Complex Magnetic Fields}

\author[R. V. E. Lovelace et al.]
{R. V. E. Lovelace$^1$\thanks{E-mail: RVL1@cornell.edu}, M. M. Romanova$^2$\thanks{E-mail: romanova@astro.cornell.edu}, G.
V. Ustyugova$^3$\thanks{E-mail: ustyugg@rambler.ru}, A. V.
Koldoba$^4$\thanks{E-mail: koldoba@rambler.ru}\\
$^1$ Departments of Astronomy and Applied and Eng. Phys. Cornell University,
Ithaca, NY 14853, USA\\
$^2$ Department of Astronomy, Cornell University, Ithaca, NY 14853, USA\\
$^3$ Keldysh Institute of Applied Mathematics, Russian Academy of
Sciences, Moscow, Russia \\
$^4$ Institute for Mathematical Modeling, Russian Academy of
Sciences, Moscow, Russia
}

\begin{document}
\maketitle \label{firstpage}

\begin{abstract}
\noindent We investigate the generation of
intrinsically asymmetric or {\it one-sided} outflows
or jets from disk accretion
onto rotating stars with complex magnetic fields
using axisymmetric (2.5D) magnetohydrodynamic simulations.
   The intrinsic magnetic field of the star is assumed to
consist of a superposition of an aligned dipole and an
aligned quadrupole in different proportions.
   The star is assumed to be rapidly rotating in
the sense that the star's magnetosphere is in the propeller regime
where strong outflows occur.
    Our simulations show that for conditions where there is
a  significant quadrupole component in
addition to the dipole component, then
a dominantly {\it one-sided} conical wind tends to form
on the side of the equatorial plane with the
larger value of the intrinsic axial magnetic field
at a given distance.
     For cases where the  quadrupole component
is absent or very small, we find that
dominantly one-sided outflows also form, but
the direction of the flow ``flip-flops'' between upward and downward on a time-scale of $\sim 30$ days for a protostar.   The average outflow will
thus be symmetrical.   In the case of a pure quadrupole field  we find
symmetric outflows in the upward and downward directions.

\end{abstract}


\section{Introduction}

There is clear evidence,  mainly from
Hubble Space Telescope (HST) observations,  of the asymmetry
between the approaching and receding jets
from a number of young stars.
The objects include the jets in HH 30 (Bacciotti et al. 1999),
RW Aur (Woitas et al. 2002), TH 28 (Coffey et al. 2004),
and LkH$\alpha$ 233  (Pererin \& Graham 2007).
  Specifically, the radial speed of the approaching
jet may differ by a factor of two from that of
the receding jet.
    For example, for RW Aur the radial redshifted
speed is $\sim 100$ km/s whereas the blueshifted
radial speed is $\sim175$ km/s.
   The mass and momentum fluxes are also
significantly different
for the approaching and receding jets in a number
of cases.
  Of course, it is possible that the observed asymmetry
of the jets could be due to say differences in the gas densities
on the two sides of the source.   Here, we investigate
the case of intrinsic asymmetry where the asymmetry of outflows is
connected with asymmetry of the star's magnetic field.

There is substantial  observational evidence that
young stars often have {\it complex} magnetic  fields consisting
of    dipole,  quadrupole, and higher order  poles possibly
misaligned with respect to each other and the rotation
axis (Donati et al. 2007a, b; 2008;  Jardine et al. 2002).
Analysis of mater flow around stars with realistic fields
have shown that a fraction
of the star's magnetic field lines are open
and may carry outflows (e.g., Gregory et al. 2006).

  A number of global 3D MHD simulations have been done of disk accretion
onto rotating stars with complex magnetic fields.
   The star's intrinsic field may be a
 superposition of aligned or misaligned dipole and quadrupole fields (Long, Romanova, \& Lovelace 2007, 2008), or a superposition of  dipole and octupole field components (Long, Romanova, Lamb 2009; Romanova et al. 2009a; Long et al. 2010).
    These simulations were focused on accretion processes.
     To study the outflows from these systems requires
a much lower coronal density than assumed in these works.
   Intermittent outflows from the disk-magnetosphere boundary have been
found in axisymmetric  simulations in cases where the star has a dipole magnetic field   and where symmetry about the equatorial plane was assumed (e.g., Goodson, Winglee, \& B\"ohm 1997;  Goodson, B\"ohm, \& Winglee 1999).

     In long-time axisymmetric (2.5D) simulations,   long-lasting outflows were obtained first in the propeller regime where the star
spins rapidly (Romanova et al. 2005; Ustyugova et al. 2006), 
and subsequently in the general case by Romanova et al. (2009b;  hereafter
R09).
    In the propeller regime the outer part of the star's magnetosphere
    - the magnetopause -
rotates more rapidly than the Keplerian rate of the accretion disk
(Lovelace, Romanova, \& Bisnovatyi-Kogan 1999).    
      Simulations show that most of the matter outflow
goes into  conical-shape winds from the inner part of the disk.
     This wind resembles the X-wind model
(Shu et al. 1994), but there are a number of important
differences discussed in R09.

    The present work investigates the nature of
outflows from a star with a {\it complex magnetic
field}. As a first step we consider superposition of
an dipole and a quadrupole field components
both aligned with the rotation axis of the star and of the disk.
    For such a configuration  the magnetic field is {\it not} symmetric
about the equatorial plane.   This can give rise to one-sided or
asymmetric  magnetically
driven outflows  (Wang, Sulkanen, \& Lovelace 1992).
          Figure 1 shows the nature of a combined
vacuum dipole plus quadrupole field components.
In this  case we
have sketched an extreme possibility where there is a
conical wind from the top side of the disk but no outflow
from the bottom side, only a funnel flow marked ff.

\begin{figure}
\centering
\includegraphics[width=3.4in]{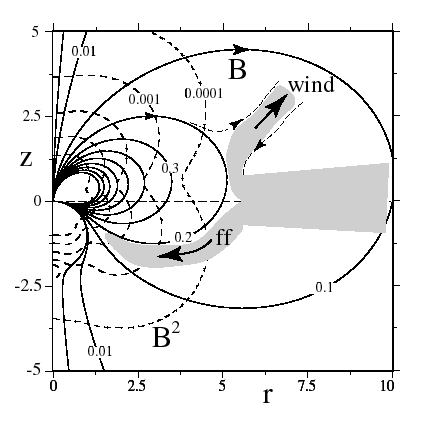}
\caption{The magnetic field lines $\Psi(r,z)=$ const
and constant magnetic pressure lines for
the case of an aligned dipole and quadrupole field
where the flux function is $\Psi =\mu_d r^2/R^3+(3/4)\mu_q z r^2/R^5$, where
$R^2=r^2+z^2$ and $\mu_d$ is the dipole moment and $\mu_q$ is the
quadrupole moment.   Roughly, $\mu_q/\mu_d$ is
 the distance
at which dipole and quadrupole fields are equal.
The funnel flow (ff) and the wind in this figure are suggested.
The dashed lines are constant values of ${\bf B}^2$.}
\label{sketch}
\end{figure}

In this paper  we present results out systematic
 MHD simulations of the formation of conical winds
for the axisymmetric dipole/quadrupole field combinations
following the approach of R09 where
conical winds are found to form at the disk-magnetosphere
boundary.  The new aspect of the present work is that the
outflows are {\it not} required to be symmetrical about
the equatorial plane.

    Section 2 of the paper describes the different aspects
of the simulation model.  Section 3 discusses the results
of the simulations
  We will quantify the asymmetry
of the velocities of the top/bottom jets as well
as the asymmetry of the mass and momentum fluxes
for different quadrupole/dipole field strengths, different
disk accretion rates, and different stellar rotation rates.
    Section 4 gives the conclusions of this work.

\begin{figure*}
\centering
\includegraphics[width=7in]{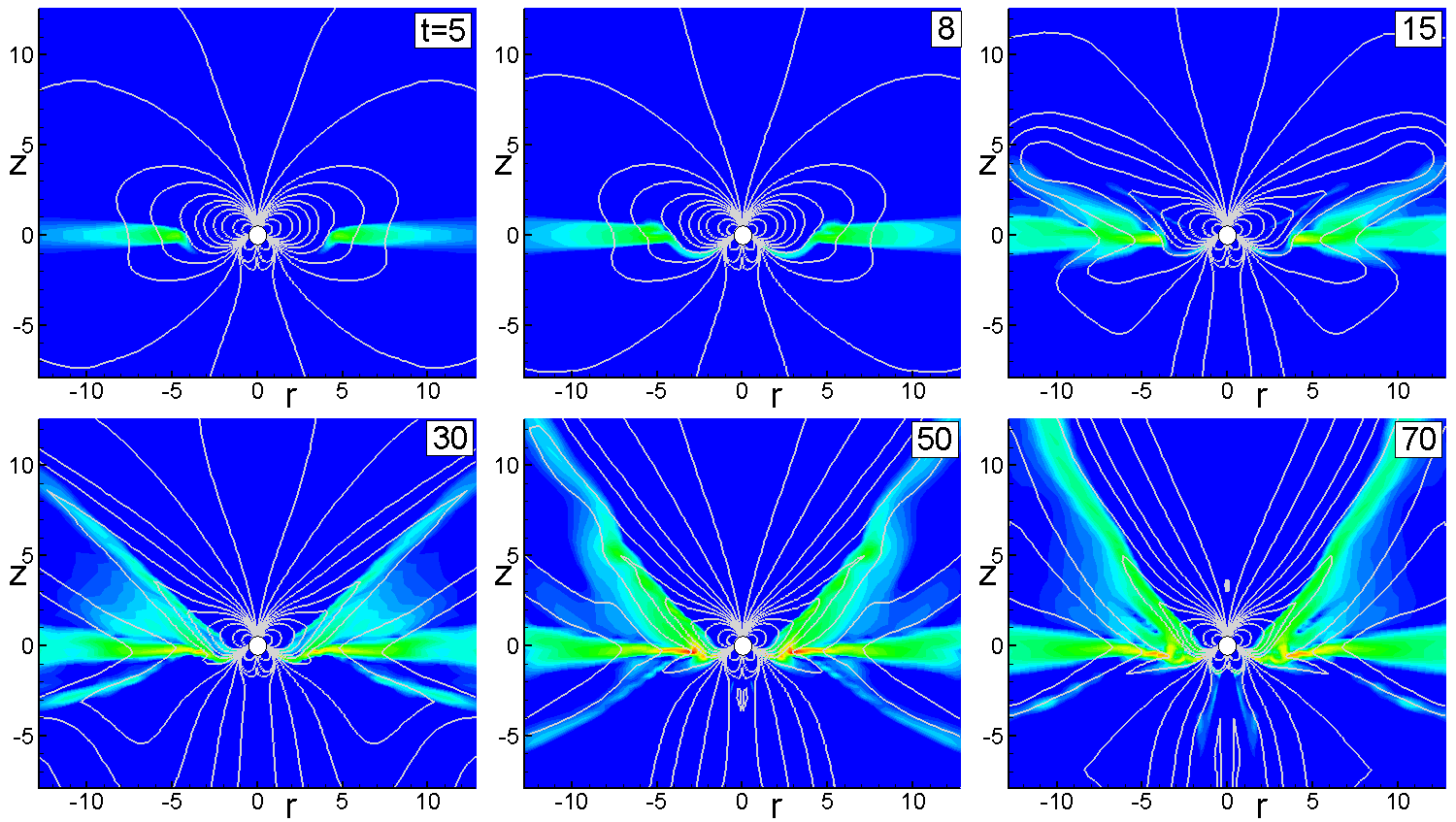}
\caption{Development of an asymmetric jet in the main case ($\tilde{\mu}_d=10$, $\tilde{\mu}_q=20$).
The color background shows the matter flux-density. 
The lines are the poloidal field lines.  The simulations are
shown at different  times $t$, which is measured in periods of rotation of
the disk at $r=1$.}
\label{q20-evol-6}
\end{figure*}

\begin{figure*}
\centering
\includegraphics[width=6.in]{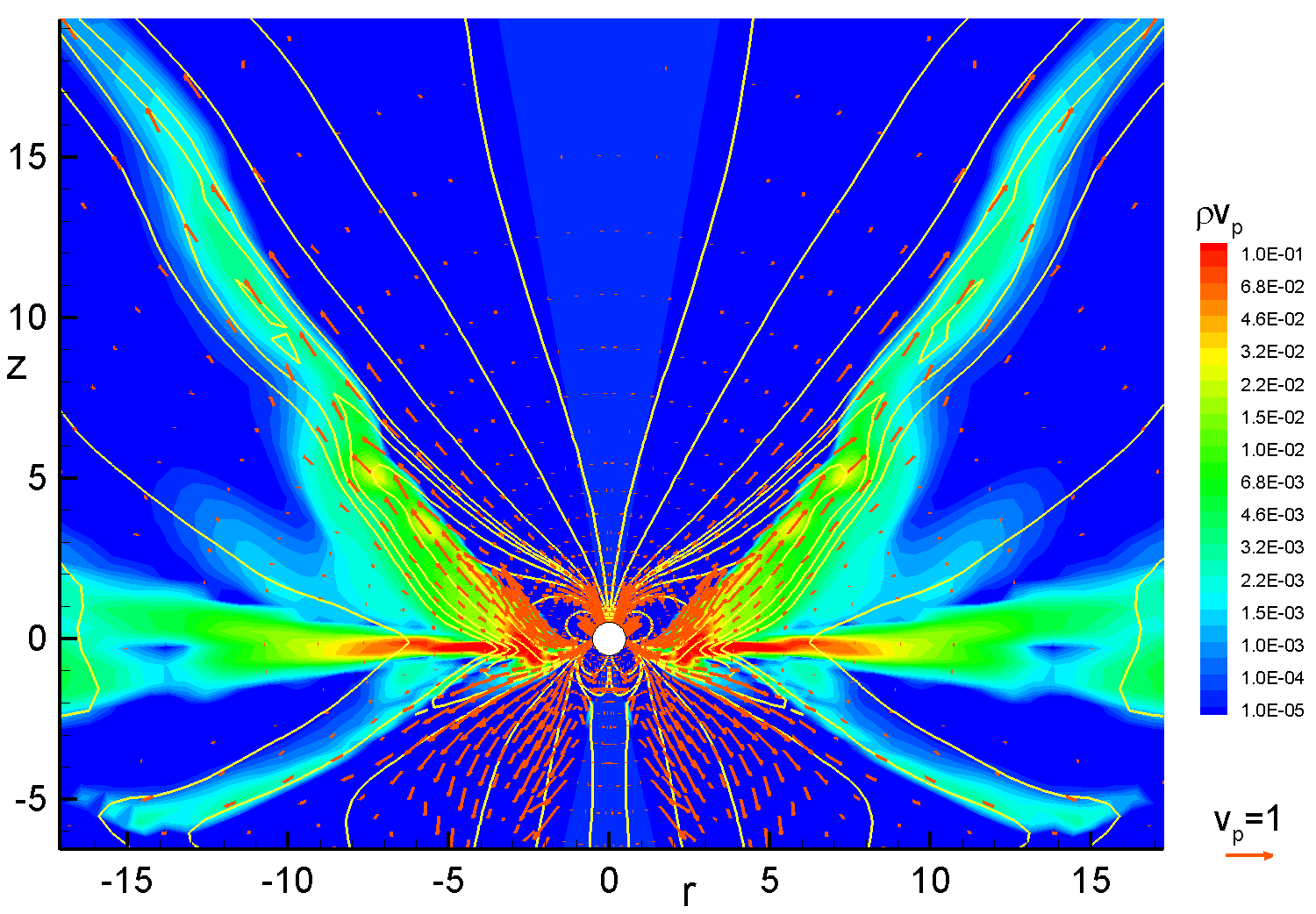}
\caption{Enlarged snapshot of the outflow
in the main case ($\tilde{\mu}_d=10$, $\tilde{\mu}_q=20$) at $t=50$.
The color background shows the matter flux-density
and the lines are the poloidal field lines.   The vectors
show the poloidal velocity.}
\label{q20-big}
\end{figure*}

\begin{figure*}
\centering
\includegraphics[width=3.in]{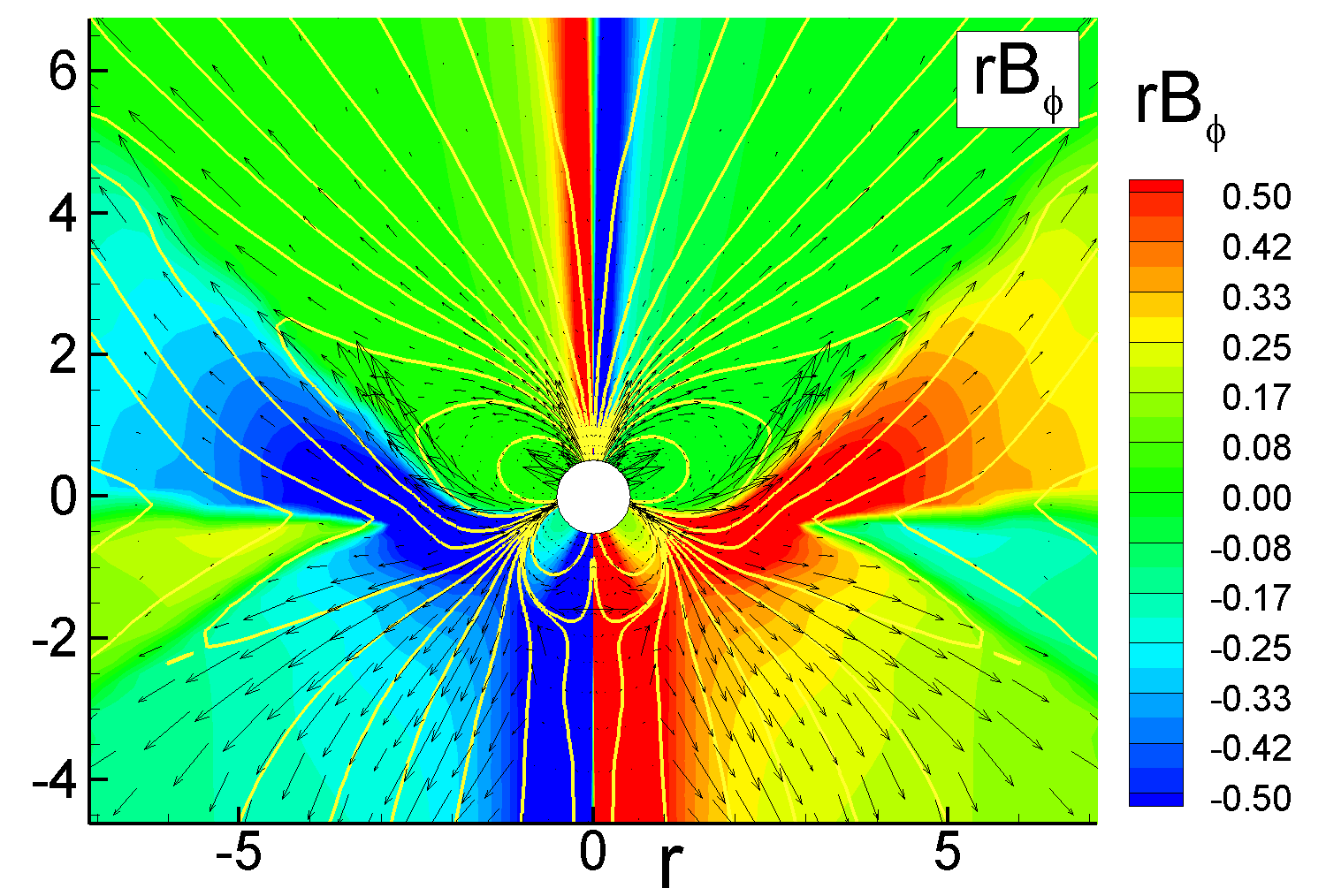}
\caption{The color background shows  $rB_\phi(r,z)$ at
$t=50$,
which is proportional to the poloidal current flow through
a circular disk of radius $r$ at a distance $z$.  The lines are poloidal field
lines and the arrows show the poloidal velocity.
For this case $\tilde{\mu}_d=10$ and $\tilde{\mu}_q=20$.}
\label{q20-phys-j}
\end{figure*}

\begin{figure*}
\centering
\includegraphics[width=3.5in]{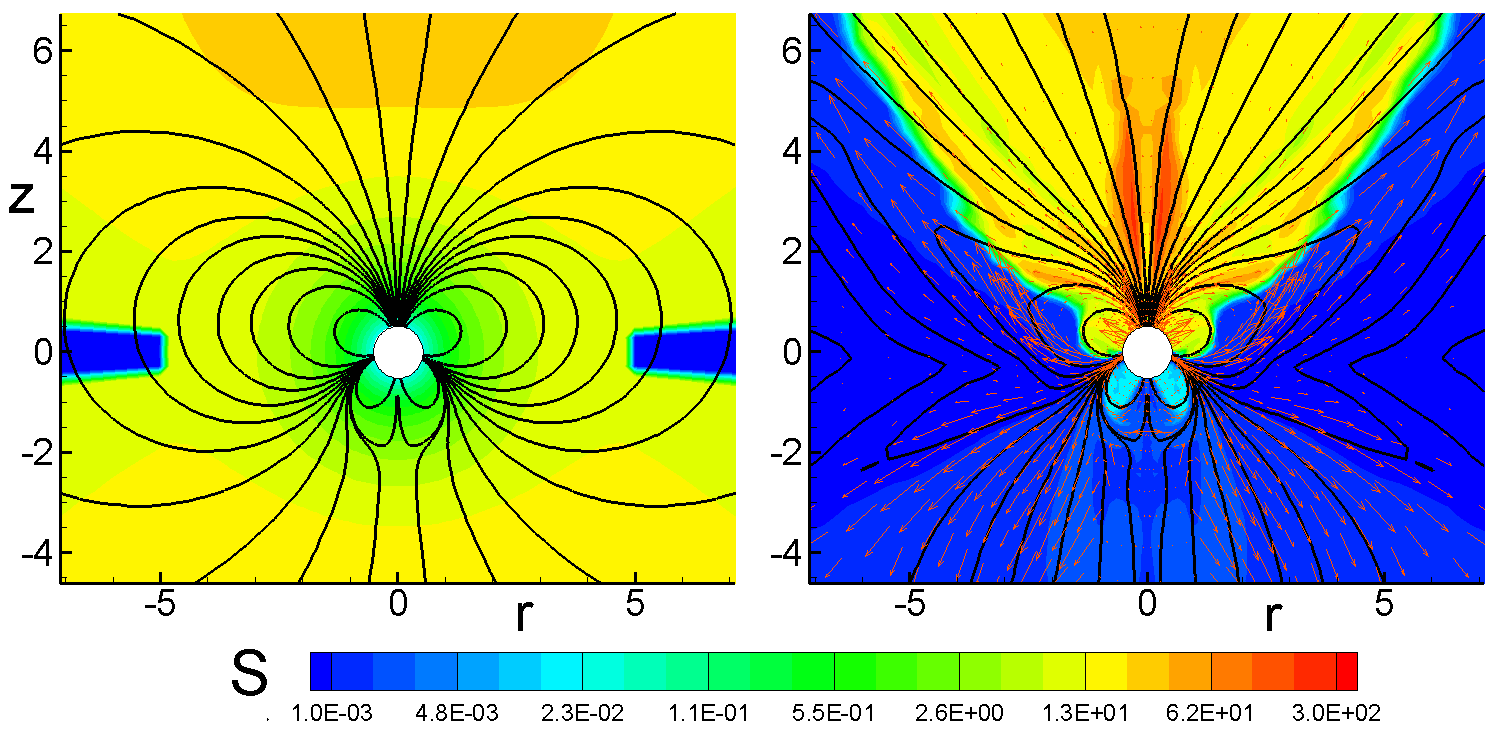}
\caption{The color background shows the distribution of
the specific entropy  $S$.   The left-hand panel shows
initial distribution
and the right-hand panel shows the distribution at $t=50$.
The lines are poloidal  field lines.
For this case $\tilde{\mu}_d=10$ and $\tilde{\mu}_q=20$.}
\label{q20-s-2}
\end{figure*}

\begin{figure*}
\centering
\includegraphics[width=7in]{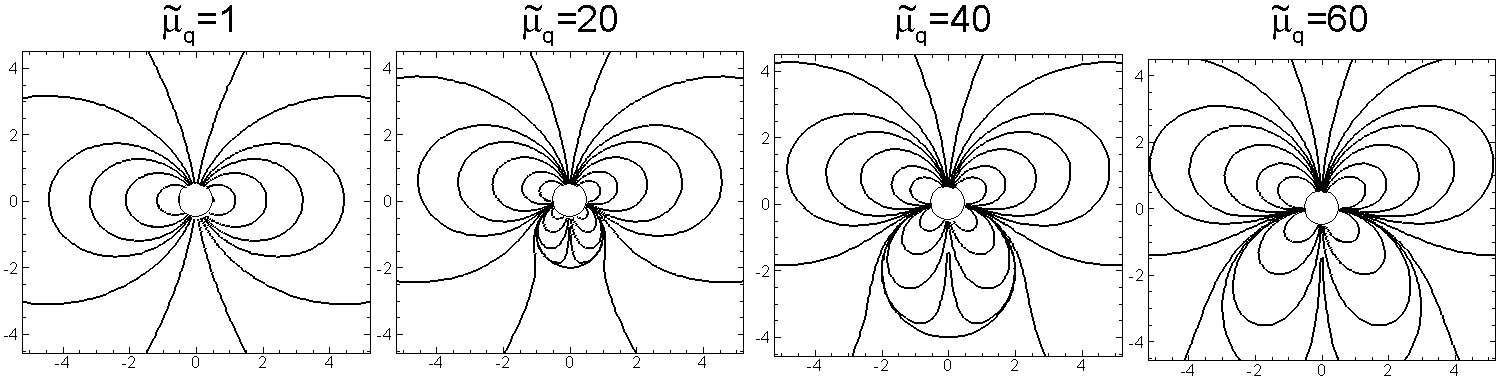}
\caption{The figure shows
the initial poloidal field lines for a fixed dipole moment $\mu_d=10$ and different
quadrupolar moments, $\mu_q=0, 20, 30, 40$.
   The field lines have the same value of the flux function
in all plots.}
\label{conf-4}
\end{figure*}

\begin{figure*}
\centering
\includegraphics[width=7in]{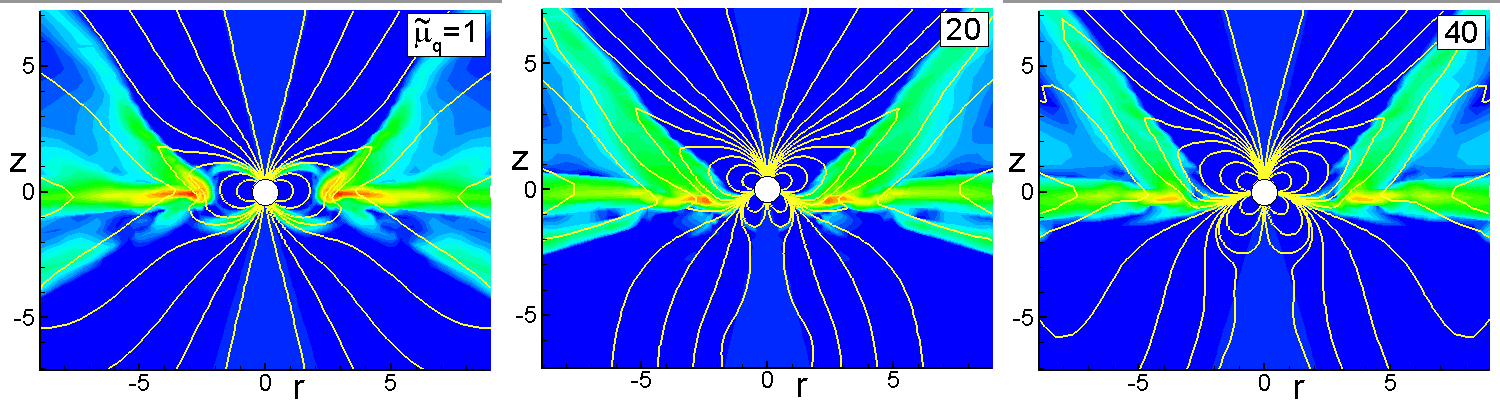}
\caption{Outflows for different quadrupole strengths $\mu_q$ but
the  same dipole component
$\mu_d=10$  at $t=50$. The color background shows the matter flux-density
distribution.   The lines are poloidal magnetic field lines.}
\label{q-dif-3}
\end{figure*}

\begin{figure*}
\centering
\includegraphics[width=3.5in]{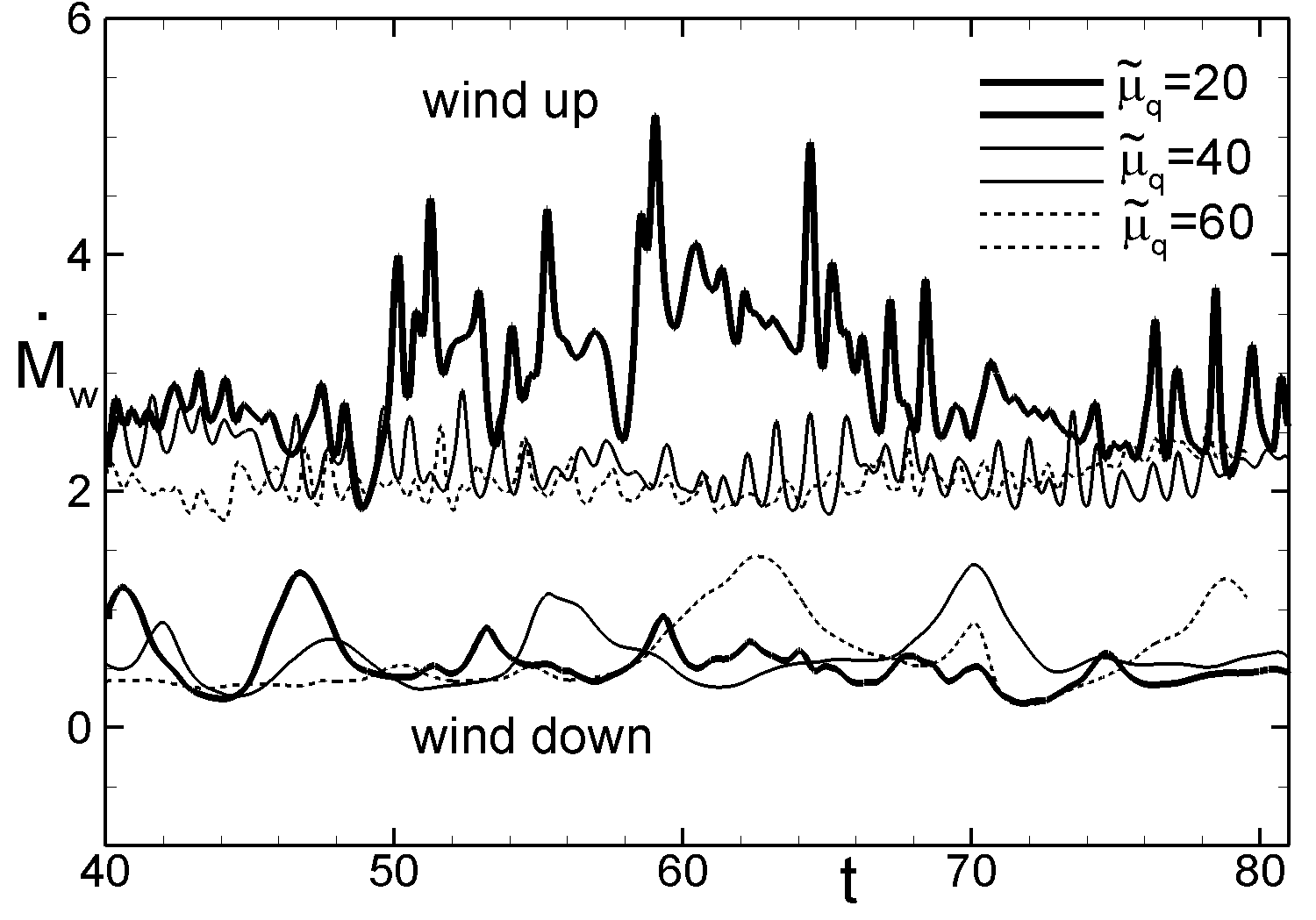}
\caption{Matter fluxes  through a hemispherical
surface of radius $r=10$ above the equatorial plane
(top set of curves) and below the equatorial plane (bottom set
of curves) for cases with $\tilde{\mu}_d=10$ and different quadrupole moments of $\tilde{\mu}_q=20, 40, 60$.}
\label{flux-q}
\end{figure*}

\section{Simulation Model}

We investigate the formation of one-sided
or asymmetric jets using a set of
axisymmetric numerical simulations.
    The arrangement of the problem is
similar to that described
by Ustyugova et al. (2006, hereafter - U06).
However, in contrast with U06,  the simulations
were performed in the {\it entire} simulation region
{\it without} assuming symmetry about the equatorial
plane.
 Below we describe the main aspects of the simulation model
and point out the new aspects of the present model.

\subsection{Basic Equations}

Outside of the disk the flow is described by the
equations of ideal MHD. Inside the disk the flow is described by
the equations of viscous, resistive MHD.
In an inertial reference frame the equations are
\begin{equation}\label{eq1}
\displaystyle{ \frac{\partial \rho}{\partial t} + {\bf \nabla}\cdot
\left( \rho
{\bf v} \right)} = 0~,
\end{equation}
\begin{equation}\label{eq2}
{\frac{\partial (\rho {\bf v})}{\partial t} + {\bf
\nabla}\cdot
{\cal T} } = \rho ~{\bf g}~,
\end{equation}
\begin{equation}\label{eq3}
{\frac{\partial {\bf B}}{\partial t} - {\bf
\nabla}\times ({\bf v} \times {\bf B}) + {\bf \nabla} \times\left(
\eta_t {\bf \nabla}\times {\bf B} \right)} = 0~,
\end{equation}
\begin{equation}\label{eq4}
{\frac{\partial (\rho S)}{\partial t} + {\bf
\nabla}\cdot ( \rho S {\bf v} )} =  Q~.\
\end{equation}
 Here, $\rho$ is the density and $S$ is the specific entropy; $\bf
v$ is the flow velocity; $\bf B$ is the magnetic field; $\cal{T}$ is the
momentum flux-density tensor; $Q$ is
the rate of change of entropy per unit volume;   and ${\bf g} = - (GM /r^{2})\hat{{\bf r}}$
is the gravitational acceleration due to the star, which has mass $M$.
    The total
mass of the disk is assumed negligible compared to $M$.
     The  plasma is considered to be an
ideal gas with adiabatic index $\gamma =5/3$, and $S=\ln(p/
\rho^{\gamma})$. We use spherical coordinates $(r, \theta, \phi)$ with  $\theta$ measured
from the symmetry axis.
     The condition for axisymmetry is $\partial /\partial
\phi =0$.
 The  equations in spherical coordinates are given in
U06.

The stress tensor $\cal T$ and the treatment of viscosity and diffusivity
are described in the Appendix of R09.
     Briefly,  both the viscosity
and the magnetic diffusivity of the disk plasma are considered
to be due to turbulent fluctuations of the velocity and the magnetic field.
        We adopt the standard hypothesis
where the  microscopic transport coefficients are  replaced by
 turbulent coefficients.
       We  use the  $\alpha$-model of Shakura and Sunyaev (1973) where the
coefficient of the turbulent kinematic viscosity $\nu_t =
\alpha_v c_s^{2}/\Omega_K$, where $c_s$ is the isothermal sound
speed and $\Omega_K(r)$ is the Keplerian angular velocity.
Similarly, the coefficient of the turbulent magnetic
diffusivity $\eta_t=\alpha_d c_s^{2}/\Omega_K$. Here,
$\alpha_v$ and $\alpha_d$ are dimensionless coefficients which
are treated as parameters of the model.

\subsection{Magnetic field of star}

We consider the superposition of aligned dipole and quadrupole field
components,
\begin{equation}
\label{e0} \mathbf{B}=\mathbf{B}_d+\mathbf{B}_q ~,
\end{equation}
where
\begin{eqnarray}
\label{e1}
{\bf B}_{d}&=&{3\mu_d({\hat{\bm\mu}}_d\cdot{\hat{\bf r}})
\hat{\bf r}\over |{\bf r}|^3}- {\mu_d\hat{\bm\mu}_d \over |{\bf r}|^3 }~,\nonumber\\
\mathbf{B}_{q}&
=&{3\mu_q(5({\hat{\bm\mu}}_q\cdot\hat{\bf r}) ^2-1)\hat{\bf r}\over 4 |{\bf r}|^4}
-{3\mu_q(\hat{\bm\mu}_q\cdot\hat{\bf r})
\hat{\bm\mu}_q\over 2 |{\bf r}|^4}~.\nonumber\\
\end{eqnarray}
Here, ${\bf B}_d$ and ${\bf B}_q$ are the magnetic fields of the  dipole and quadrupole components with 
$\mu_d$ and $\mu_q$  the magnetic moments.
  Also,  $ \hat{\bf r}$, $\hat{\bm\mu}_d$ and $\hat{\bm\mu}_q$
are  unit vectors for the position and the direction
of the dipole and quadrupole moments, respectively.  For
the considered conditions $\hat{\bm\mu}_d$ and $\hat{\bm\mu}_q$ are in the $z-$direction.   The combined dipole/quadrupole field can be expressed
in terms of the flux function  $\Psi =\mu_d r^2/R^3+(3/4)\mu_q z r^2/R^5$,
where the field lines correspond to $\Psi(r,z)=$ const.
Here, we briefly use cylindrical coordinates $(r,~\phi,~z)$,
with $R^2=r^2+z^2$,  $B_r =-(1/r)\partial \Psi/\partial z$,
and $B_z= (1/r)\partial \Psi/\partial r$.

\subsection{Reference Units}

The MHD equations are solved in
dimensionless form so that the results can be readily applied
to different accreting stars (see \S 7).
   We take the reference
mass $M_0$ to be the mass $M$ of the star.
  The reference radius is taken to be {\it twice}   the radius
of the star, $R_0=2\times R_*$.
      The reference velocity is $v_0=(GM/R_0)^{1/2}$,
 and the reference angular velocity $\Omega_0=1/t_0$.
    We measure time
in units of $P_0=2\pi t_0$, which is the Keplerian
rotation period of the disk at $r=R_0$.
In the plots we  use the dimensionless time $T=t/P_0$.

The dimensionless dipole and quadrupole magnetic moments are 
\begin{equation} 
\tilde\mu_d\equiv {\mu_d \over B_0 R_0^3}~,\quad \tilde\mu_q\equiv {\mu_q\over B_0 R_0^4}~,
\end{equation}
where $B_0$ is the reference magnetic field.
      Taking into account that 
$\mu_d=B_{d\star}R_\star^3=B_0R_0^3\tilde\mu_d$, we 
find $B_0=B_{d\star}(R_\star/R_0)^3/\tilde\mu_d$, where $B_{d\star}$ is the equatorial dipole magnetic field  strength on
the surface of the star.

The reference density is taken to
be  $\rho_0 = B_0^{2}/v_0^{2}$.
The reference pressure is $p_0=B_0^{2}$.
The reference temperature is
$T_0=p_0/{\cal R} \rho_0 = v_0^{2}/{\cal R}$, where ${\cal R}$ is the gas constant.
   The reference accretion rate is $\dot M_0
= \rho_0 v_0 R_0^{2}$.
   The reference energy flux is $\dot
E_0=\dot M_0 v_0^{2}$.
The reference angular momentum flux is $\dot L_0=\dot M_0 v_0
R_0$.

The reference units are defined in such
a way that the dimensionless MHD equations
have the same form as the dimensional ones, equations (1)-(4) (for such dimensionalization
we put $GM=1$ and ${\cal R}=1$).
Table 1 shows examples of reference variables for
different stars.
We solve the MHD equations (1)-(4) using normalized variables:
$\tilde\rho=\rho/\rho_0$, $\tilde v=v/v_0$, $\tilde B_d= B_d/B_0$, $\tilde B_q= B_q/B_0$ etc.
Most of the plots show the normalized variables (with the tildes
implicit). To obtain dimensional values one needs to multiply
values from the plots by the corresponding  reference values from
Table 1.

\begin{table*}
\begin{tabular}{l@{\extracolsep{0.2em}}l@{}lllll}

\hline
&                                              & Protostars      & CTTSs        & Brown dwarfs   & White dwarfs        & Neutron stars    \\
\hline

\multicolumn{2}{l}{$M(M_\odot)$}                  & 0.8            & 0.8            & 0.056             & 1                   & 1.4       \\
\multicolumn{2}{l}{$R_\star$}                     & $2R_\odot$     & $2R_\odot$     & $0.1R_\odot$       & 5000 km             & 10 km     \\
\multicolumn{2}{l}{$R_0$ (cm)}                    & $2.8\e{11}$    & $2.8\e{11}$    & $1.4\e{10}$      & $1.0\e9$            & $2\e6$    \\
\multicolumn{2}{l}{$v_0$ (cm s$^{-1}$)}           & $1.95\e7$      & $1.95\e7$      & $1.6\e7$          & $3.6\e8$   & $9.7\e{9}$ \\
\multicolumn{2}{l}{$P_0$}                         & $1.04$ days    & $1.04$ days    & $0.05$ days       & $17.2$ s            & $1.3$ ms   \\
\multicolumn{2}{l}{$B_{d\star}$ (G)}              & $3.0\e3$      & $10^{3}$      & $2.0\e{3}$            & $10^{6}$            & $10^{9}$    \\
\multicolumn{2}{l}{$B_0$ (G)}                     & 37.5           & 12.5          & 25.0                & $1.2\e4$            & $1.2\e{7}$  \\
\multicolumn{2}{l}{$\rho_0$ (g cm$^{-3}$)}         & $3.7\e{-12}$  & $4.1\e{-13}$  & $1.4\e{-12}$        & $1.2\e{-9}$         & $1.7\e{-6}$  \\
\multicolumn{2}{l}{$\dot M_0$($M_\odot$yr$^{-1}$)}  & $1.8\e{-7}$  & $2.0\e{-8}$   & $1.8\e{-10}$      & $1.3\e{-8}$         & $2.0\e{-9}$  \\
\multicolumn{2}{l}{$\dot E_0$ (erg s$^{-1}$)}       & $2.1\e{33}$  & $2.4\e{32}$   & $2.5\e{30}$      & $5.7\e{34}$         & $6.0\e{36}$  \\
\multicolumn{2}{l}{$\dot L_0$ (erg s$^{-1}$)}       & $3.1\e{37}$  & $3.4\e{36}$   & $1.7\e{33}$      & $1.6\e{35}$         & $1.2\e{33}$  \\
\multicolumn{2}{l}{$T_d$ (K)}                       & $2293$       & $4586$        & $5274$            & $1.6\e{6}$          & $1.1\e{9}$  \\
\multicolumn{2}{l}{$T_c$ (K)}                       & $2.3\e{6}$   & $4.6\e{6}$    & $5.3\e{6}$        & $8.0\e{8}$          & $5.6\e{11}$  \\
\hline
\end{tabular}
\caption{Reference values for different types of stars. We choose the mass $M$, radius $R_\star$,
equatorial dipole magnetic field $B_{d\star}$ of the star and derive the other reference values  for the case $\tilde{\mu}_d=10$ (see \S 2.3).
To apply the simulation results to a particular star one needs to multiply the dimensionless values
from the plots by the reference values from this table.} \label{tab:refval}
\end{table*}

\subsection{Initial and Boundary Conditions}

The initial and boundary conditions
are the same as those used in U06.
Here, we summarize these conditions.

{\it Initial Conditions}. 
A star of mass $M$ is located
at the origin of the coordinate system.
   A cold disk and hot corona are initialized in the simulation region.      
The disk is of
low-temperature  $T_d$ and 
high-density $\rho_d$. The corona is of 
high-temperature $T_c\gg T_d$, and
low-density $\rho_c \ll \rho_d$ and it fills all other space but the disk.
   The disk extends inward to a radius $r_d=5$ and rotates with 
Keplerian angular velocity $\omega \approx \Omega_K$.
In reality, it is slightly sub-Keplerian,
$\Omega(\theta=\pi/2)=\kappa \Omega_K$ ($\kappa = 1-0.003$), due to which the density and
pressure decrease towards the periphery.
Initially, at
any cylindrical radius $r$ from
the rotation axis, we rotate
the corona and the disk
at the same angular rate.
    This avoids a jump
discontinuity of the angular velocity
of the plasma at the boundary
between the disk and the
corona.
      Inside the cylinder
$r\leq r_d$, the matter of the corona rotates rigidly with angular velocity
$\Omega(r_d)=\kappa (GM /r_d^{3})^{1/2}$.
      For a gradual start-up we change the angular velocity of the
star from its initial value $\Omega(r_d) =
5^{-3/2} \approx 0.09$ ($r_d=5$) to a final value of $\Omega_*=1$ over the course of three
Keplerian rotation periods at $r=1$.

In most simulation runs we fix the 
dipole moment of the star as $\tilde\mu_d=10$ and vary the quadrupole
moment as $\tilde{\mu}_q=0, 1, 10, 20, 30, 40, 60$. We also have test cases of $\tilde{\mu}_d=0$ and $\tilde{\mu}_q=20, 60$.
The angular velocity of the star in the propeller regime is $\Omega_*=1$ 
and this corresponds to the corotation
radius of $r_{cor}=1$. The initial density in the disk at the fiducial point (at the
inner edge) is $\rho_d=1$, initial density in the corona $\rho_c=0.0003$.  The gas in the corona is hot with
initial temperature
 $T_c=1$ and the disk is cold with temperature $T_d=(\rho_c/\rho_d) T_c = 3\times 10^{-4}$.
 There is initial  pressure equilibrium at the
disk-corona boundary.

The coefficients of viscosity and diffusivity are taken to be
$\alpha_v=0.3$ and $\alpha_d=0.1$ (as in R09).

{\it Boundary conditions}.
The boundary conditions at the {inner boundary $r=R_{in}$} are the following:
The frozen-in condition is applied to the poloidal component $\bf{B}_p$  of
the field, such that $B_r$  is fixed while $B_\theta$ and $B_\phi$ obey
 ``free" boundary conditions,  $\partial B_\theta/\partial r=0$
 and $\partial B_\phi/\partial r=0$.
The density, pressure, and entropy  also have free boundary conditions,
$\partial (...)/\partial r=0$.
The velocity components are calculated using  free boundary conditions.
    Then, the velocity vector is adjusted to be parallel to the magnetic field vector
in the coordinate system rotating with a star. Matter always flows inward at
the star's surface.
   Outflow of a wind from the stellar surface is not considered in this work.
The boundary conditions at the {\it external boundary $r=R_{out}$}
in the {\it coronal region}  $0<\theta <\theta_{d1}$ and 
$\theta_{d2} <\theta \leq \pi$ are
 free for all hydrodynamic variables.
    Here, $\theta_{d1}$ corresponds to the top surface of the
and $\theta_{d2}$ to the bottom surface.
  We prevent
 matter from flowing into the simulation region from this part of the
boundary. We solve the transport equation for the flux function $\Psi$
so that the magnetic flux flows out of the region together with matter.
   If the matter has a tendency to flow back in, then we fix $\Psi$.
    In the disk region, restricted by two values of $\theta$ 
    ($\theta_{d1}<\theta <\theta_{d2}$), we fix the density at
$\rho=\rho_d$, and establish a slightly sub-Keplerian velocity, $\Omega_d=\kappa\Omega(r_d)$,
where $\kappa = 1 - 0.003$ so that matter flows into the simulation region through the boundary.
The inflowing matter has a fixed magnetic flux which is very small because $R_{out} \gg R_{in}$.

The system of MHD equations (\ref{eq1}-\ref{eq4})
was integrated numerically using the Godunov-type numerical scheme
(see Appendix of R09).
The simulations were done
in the region $R_{in} \leq r \leq R_{out}$, $0\leq \theta
\leq \pi$. The grid is uniform in the $\theta$-direction.
    The size steps in the radial direction were chosen
so that the poloidal-plane cells were
curvilinear rectangles with approximately
equal sides.
      A typical region for investigation of asymmetric winds was
$1 \leq r \leq 40$,
with grid resolution $N_r\times N_\theta = 104\times 80$ cells.
The simulation domain has $13$ slices in the radial direction and $10$ slices in $\theta$ direction.
Each simulation run takes $4-12$ days on $130$ processors of the
NASA high-performance facilities.
Cases with stronger quadrupole component require longer simulations.
   The simulation time increases with increase
of the quadrupole moment, and hence the longest
simulations are those at $\tilde{\mu}_q=60$.
Test runs were also performed at the lower grid of $80\times 60$ and higher grid
of $160\times 120$ which show similar results (approximately the
 same matter flux onto the star and into the winds),
 though the latter grids requires
more computer resources. Simulations with the grid of $80\times 60$ were used
for  a number of exploratory runs.

\section{Results}

\subsection{Properties of one-sided outflows}

Here, we chose one case with intermediate parameters: $\mu_d=10$, $\mu_q=20$, call it ``the main case", and show it in greater detail compared with other cases.
Fig \ref{q20-evol-6} shows formation of asymmetric outflows. One can see that initially,
at $t=5-8$, matter start to accrete onto the star along the shortest path, towards the quadrupolar belt,
which is below the equator. Later, at $t\approx 15$, the disk matter diffused through the
external closed field lines of the top and bottom parts of the magnetosphere, these lines
inflated, and the conical-type wind start to blow along these field lines. Most of matter outflows from the top side of the disk, where the magnetic field
is stronger.  Outflows are  episodic but
quasi-steady on average.    The episodic nature is connected with
accumulation  of matter near the closed magnetosphere, diffusion through closed field lines, inflation of these lines, outburst to the wind and enhanced accretion onto the star (see also
Goodson \& Winglee 1997; U06; R09).
Fig \ref{q20-big} shows larger view of typical, well-developed outflows at $t=50$.
One can see that outflows are more powerful
on the top part, and much less powerful on the bottom.

    The mechanism of the outflow formation is similar to that 
in cases of conical winds (R09): 
  The magnetic flux of the star is pushed towards the star
and compressed by the disk so that the poloidal field lines are always inclined relative to the disk.
In addition, the rotating magnetosphere acts to 
make the field lines to rotate with a super-Keplerian velocity.
  These field lines thread the corona which rotates more slowly.
Consequently there is a strong   
winding-up of the field lines just above the disk.
    Hence a strong magnetic force appears, $F_m = - \nabla (r^2 B_\phi^2)$, which drives matter down the gradient
of the magnetic pressure (Lovelace, Berk \& Contopoulos 1991). 
      Fig \ref{q20-phys-j} shows
the poloidal current through a disk of radius $r$, $I_p \propto r B_\phi$.
One can see that matter flows to the conical-type winds and is driven by the magnetic force.
  Ohmic heating is included in our code, 
  but it does not have a significant role in the jet launching.
Figure \ref{q20-s-2} shows the initial distribution of the specific 
entropy and that at $t=50$. 
  One can see that the low-entropy (cold)
matter from the disk flows to the winds above and below the disk. 
On the top side of the disk
this cold matter pushes the hot coronal gas
towards the axis.   On  the bottom side of the disk, the hot coronal gas is pushed away and is gradually replaced by  cold gas from the disk.

\subsection{One-sided outflows for different quadrupole moments}

In this section we compare asymmetric outflows obtained in a set of simulations
where we fixed the dipole component of the field at $\tilde{\mu}_d=10$ and varied the
quadrupolar component from very small up to very large values: $\tilde{\mu}_q=1, 10, 20, 30, 40, 60$. Figure \ref{conf-4} shows
an initial magnetic field distribution in a number of cases.   
Figure \ref{conf-4} shows that at $\tilde{\mu}_q=1$ the quadrupole component is very small and the magnetic field
is almost pure dipole field, while for $\tilde{\mu}_q=40,~ 60$
 the quadrupole field dominates.
Figure \ref{q-dif-3} shows asymmetric outflows at $t=50$. One can see that in all cases most of the matter outflows above the disk, to the side where the intrinsic
magnetic field is stronger.   At the same time there
is a much weaker matter flux from the other side  of the disk.

We calculated the matter fluxes to outflows
through a spherical surface of radius $r=10$
\begin{equation}\label{eq8}
\dot M = \int d {\bf S}\cdot \rho~{\bf v}_p~,
\end{equation}
where $d {\bf S}$ is the surface area element directed outward.
Figure \ref{flux-q} shows matter fluxes above and below the disk for cases
with different $\tilde{\mu}_q$. One can see that the main, upward outflows,
have similar matter fluxes for $\tilde{\mu}_q=40$ and $\tilde{\mu}_q=60$, while at lower
quadrupole moment $\tilde{\mu}_q=20$ matter flux is higher.  
  The matter fluxes of the downward outflows
(bottom set of curves in Fig. \ref{flux-q}) are similar for all three cases and are $4-8$ times smaller
than the main, upward outflows.

\subsection{Comparison of magnetic moments}

Below, we discuss the relative strengths of the dipole and quadrupole
components of the field. 
     For this it is useful to consider
the dipole and quadrupole field components in the equatorial plane,  
$B_{dz}=-\mu_d/r^3$ and $B_{qr}=-3\mu_q/4 r^4$.
    The radius at which  $|B_{dz}|=|B_{qr}|$  is
\begin{equation}
r_{eq}=\frac{3}{4}\frac{\mu_q}{\mu_d} ~,\quad {\rm or}\quad
r_{eq}=\frac{3}{4}\frac{\tilde\mu_q}{\tilde\mu_d} R_0 = \frac{3}{2}\frac{\tilde\mu_q}{\tilde\mu_d} R_\star~, 
\end{equation}
where we took into account that $R_0=2 R_\star$. 
    We can estimate this radius for all cases above where
$\tilde{\mu}_d=10$ and  quadrupolar 
moments, $\tilde{\mu}_q=1, 10, 20, 30, 40, 60$.
We obtain: $\tilde r_{eq}=(0.15, 1.5, 3.0, 4.5, 6.0, 9.0) R_\star$. 
Hence, the quadrupole component is dynamically important in all cases, 
except $\tilde\mu_q=1$,
and is expected to influence, e.g., the matter flow around the star at $r > r_{eq}$  
(Long et al. 2009, 2010; Romanova et al. 2009).
However, we see that the quadrupole component determines  the ``one-sideness"
and the direction of the matter outflow
even for very small values, such as  of $\tilde\mu_q=1$.

\subsection{Symmetric Outflows for a pure quadrupole field}

We performed simulations of outflows in the case of a pure quadrupole field with  $\tilde\mu_q=40, 60$  ($\tilde\mu_d=0$).
Figure \ref{quadr} shows simulation results at $t=50$ for case with
 $\tilde\mu_q=60$.  In this  case the outflows are almost symmetric 
about the equatorial plane.  However, in a test case with very small dipole magnetic field, 
$\tilde\mu_d=1$, we found that the flow is one-sided and 
matter flows towards the direction of the larger intrinsic
magnetic field.

\subsection{``Flip-flop" of Outflows for a pure dipole field}

\begin{figure*}
\centering
\includegraphics[width=7.0in]{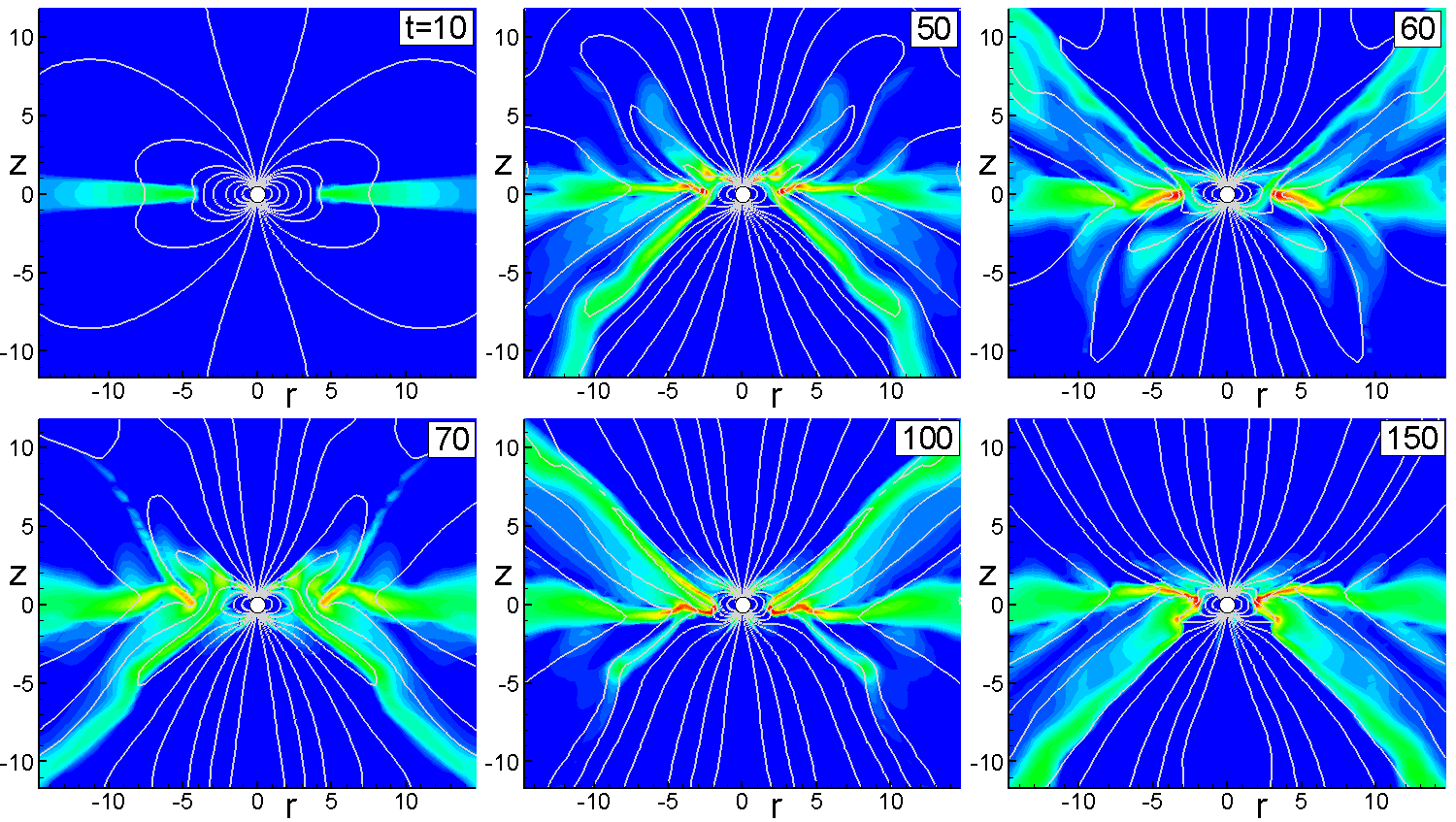}
\caption{"Flip-flop"of outflows in case of a pure dipole field ($\tilde{\mu}_q=0$, $\tilde{\mu}_d=10$.
Color background shows the matter flux
distribution, and lines are the magnetic field lines.}
\label{q0-6}
\end{figure*}

   For a pure dipole field, the
outflows behaved in an unexpected way.
In past simulations the MHD equations were solved
only in the upper half space and it was
{\it assumed} that the full flow was given
by reflecting the top flow about the equatorial plane
 (U06; R09). 
      However, in the present work
we calculate the disk and outflows in the entire space
above and below the equatorial plane.
      We discover that 
during a brief initial time, $t\lesssim 15$,
the outflows are symmetric about the equatorial plane
(see fig. \ref{q0-6}, top left panels). 
   However, later the outflows become strongly asymmetric with the
direction of the main flow downward, which is an opposite to that of  cases with quadrupolar component. 
     Later, the main outflow changed its direction 
and matter started to flow upward. 
     Later, the direction changed again
(see fig. \ref{q0-6}). 
     The matter fluxes calculated through  hemispherical
surfaces with radius $r=10$
above and below the disk also show this ``flip-flop" behavior. 
The time-scale between the  reversal of the direction of outflows is about 30 rotation periods.  For CTTSs this is about 30 days.

\begin{figure*}
\centering
\includegraphics[width=3.5in]{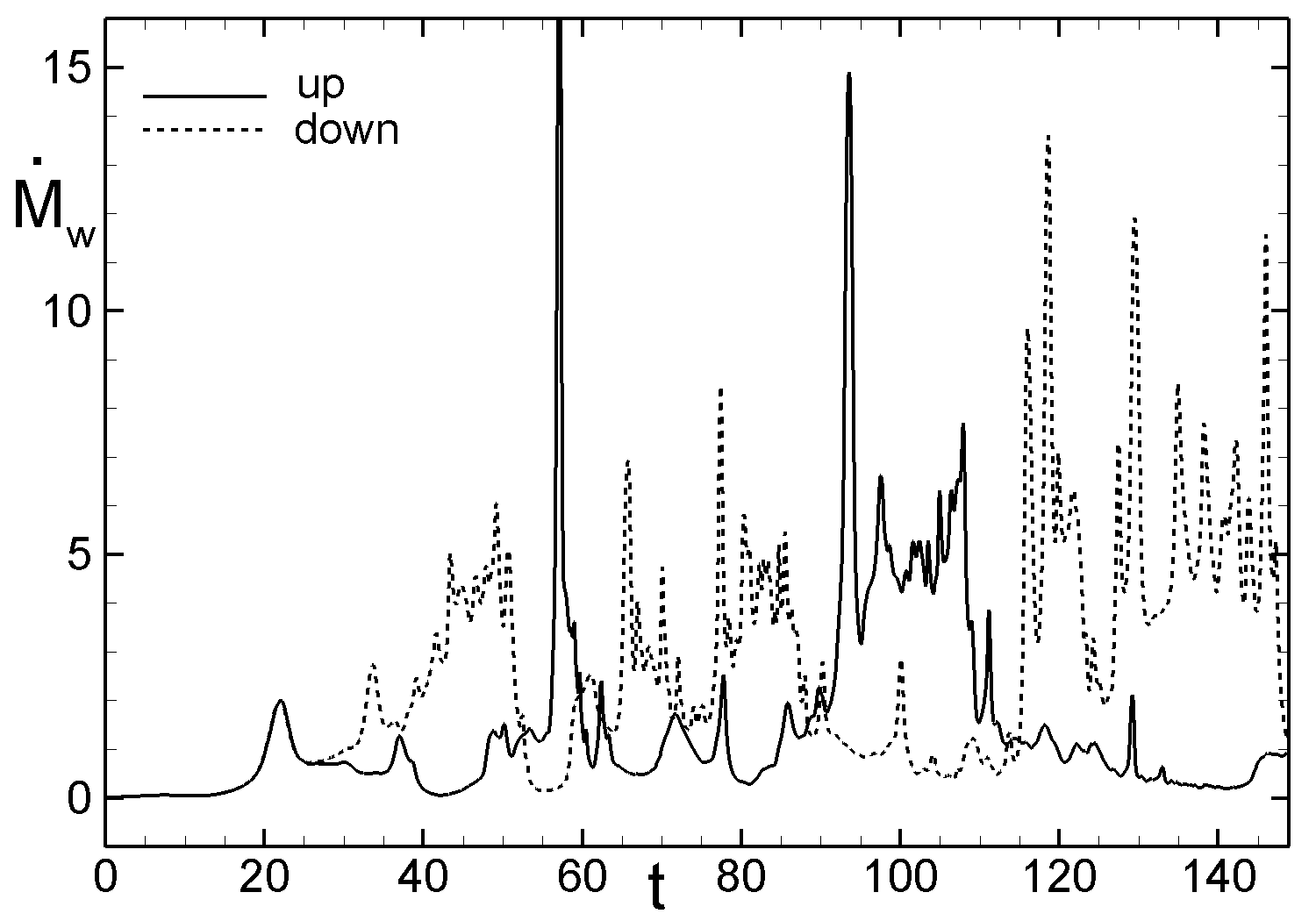}
\caption{Matter fluxes calculated through the spherical radius $r=10$ above the equatorial plane
(solid lines) and below the equatorial plane (dashed lines) for the case of pure dipole field, $\tilde{\mu}_d=10$ and $\tilde{\mu}_q=0$.}
\label{flux-0}
\end{figure*}


\begin{figure*}
\centering
\includegraphics[width=3.5in]{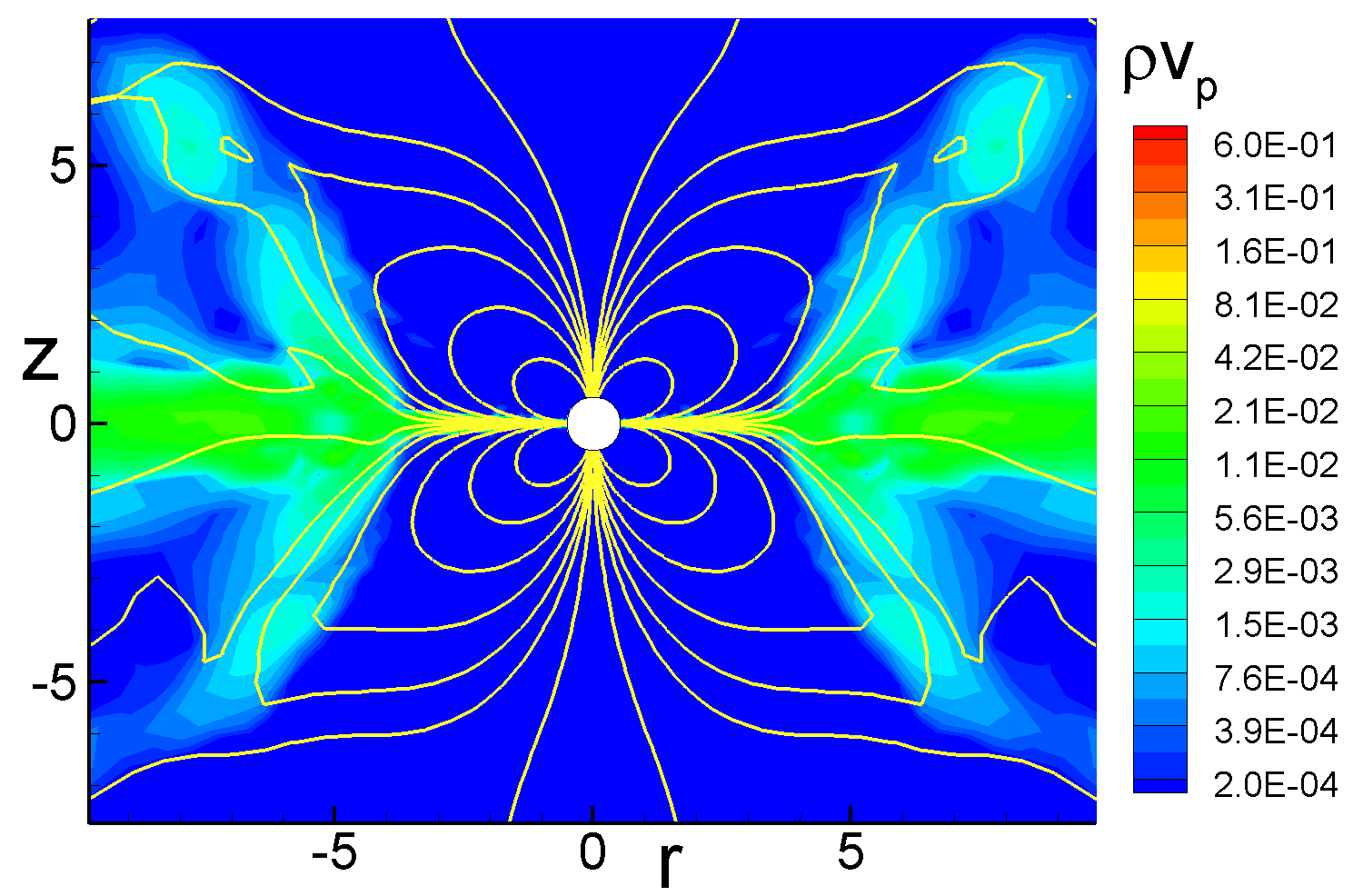}
\caption{Matter flux-density for the case of a pure quadrupole field
with $\tilde{\mu}_q =60$.  The upward and downward outflows are
approximately equal. }
\label{quadr}
\end{figure*}


\section{Conclusions}

We performed axisymmetric MHD simulations 
of  disk accretion onto rotating magnetized stars. 
   The star's intrinsic magnetic field was
assumed to consist of a superposition of aligned dipole and quadrupole
components.    This field configuration is {\it not} in general
symmetric about the equatorial plane.   Thus the calculations
must be done in both the upper and lower half-spaces.
The star was assumed to be rapidly
rotating with the magnetosphere in the propeller
regime (U06).   
     The ratio of the turbulent viscosity to the turbulent magnetic
diffusivity in the disk was considered to be larger than unity so
that the magnetic field threading the disk is advected inward (U06, R09).
The main findings are
the following:

1.   For cases with both
dipole and quadrupole components, a one-sided conical wind forms
and persistently blows in one direction.  The favored direction
is that of larger magnitude intrinsic axial magnetic field.
    Much weaker outflows form on the opposite side of the disk.

2.   For the case of a pure dipole field,  the outflows are also one-sided, but the outflow direction alternates or ``flip-flops'' on a time-scale of about
$30$d for a T Tauri star.  If the quadrupole component is small, $\tilde{\mu}_q=1$, the behavior is similar to the case of $\tilde{\mu}_q=0$.

3. For the case of a pure quadrupole field, symmetric outflows form. However, the presence of even very small dipole component leads to one-sided outflows.

     Note, that outflows from a T Tauri star may 
change direction due to  variations of the star's magnetic field
on a time-scale of  months (Smirnov et al. 2004).
     If the complex field of a star is determined by 
dynamo processes inside the star, then one or 
another hemisphere may have the stronger axial
magnetic field which determines the direction of the outflow.

    An intrinsic asymmetry of the outflows or jets will give a
net force on the protostar by analogy with the previously
analyzed case of asymmetric jets from magnetized black-hole
disks (Wang et al.  1992; Tsygan 2007; Kornreich \& Lovelace 2008).
   If the outflow is one-sided for a long time $T$ with asymptotic velocity
$V_j$ and mass flux $\dot{M}_j$, then the velocity imparted
to the star is $\Delta v_* = \dot{M}_j V_j T/M_*$.   For example,
for a T Tauri star with $\dot{M}_j =10^{-8}M_\odot$yr$^{-1}$,
$V_j=3\times 10^7$cm s$^{-1}$, $T=10^5$ yr, and
$M_*=M_\odot$, we find $\Delta v_* = 3\times 10^4 $cm s$^{-1}$
which is probably undetectable.   For the case of
an almost pure dipole field the frequent ``flip-flops'' of the
outflow direction will cause the  star to random
walk but the net displacement and velocity are very small.

       In addition to the intrinsic stellar magnetic field
considered here, the accretion disk can advect inward external
(e.g.,  interstellar) magnetic flux because of the disk's highly
conducting (non-turbulent) surface layers (Bisnovatyi-Kogan
\& Lovelace 2007; Rothstein \& Lovelace 2008; Lovelace,
Rothstein, \& Bisnovatyi-Kogan 2009).    The combination
of the advected field and the intrinsic field of the star  can give
rise to a complex field structure near the star's
magnetopause which produces  asymmetric or one sided outflows. 
  The field of the star may  be dynamo generated with a complex
time-dependent structure (e.g., von Rekowski \& Brandenburg 2006).
  In the case of disk accretion to a black hole, a large-scale asymmetric
magnetic field close to the black hole can arise from advection
of external flux due to the conducting surface layers of the disk
or it may arise from dynamo processes in the disk which generate
both dipole and quadrupole field components
(e.g., Pariev \& Colgate 2007;  Pariev, Colgate, \& Finn 2007).
Dynamo processes may also be important in the disks
of accreting stars.

\section*{Acknowledgments}

We thank F. Bacciotti
 for a valuable discussion.
This work  was supported in
part by NASA grants NNX08AH25G
and NNX10AF63G
and by NSF grant AST-0807129.
MMR thanks NASA for use of
the NASA High Performance Computing 
Facilities. AVK and GVU were supported in
part by grant RFBR 09-02-00502a, Program 4 of RAS.

\end{document}

%% file: shortcuts.tex
\newcommand{\e}[1]{\times 10^{#1}}                        





%% file: OneSided.bbl
\begin{thebibliography}{}



\bibitem{} Bacciotti, F., Eisloffel, J., \& Ray, T.P. 1999,
A\&A, 350, 917

\bibitem{} Bisnovatyi-Kogan, G.S., \& Lovelace, R.V.E. 2007,
ApJ, 667, L167

\bibitem{} Coffey, D., Bacciotti, F., Woitas, J., Ray, T.P., \&
Eisl\"offel, J. 2004, ApJ, 604, 758

\bibitem{}   Donati J.-F. Jardine, M.M., Petit, P., Morin, J.,
Bouvier, J., Cameron, A.C., Delfossse, X., Dintrans, B.,
Dobler, W., Dougados, C., Ferreira, J., Forveille, T.,
Gregory, S.G., Harries, T., Hussain, G.A.J.,
M\'enard, F., \& Faletou, F. 2007a, in  ASP Conf. Ser.,
Proc. 14th Meeting on Cool Stars, Stellar Systems and the Sun. Astron. Soc. Pac.  van Belle G., ed., (astro-ph/0702.0159)


\bibitem{} Donati, J.-F., Jardine, M. M., Gregory, S. G., Petit, P., Bouvier, J.,
Dougados, C., M\'enard, F., Cameron, A. C., Harries, T. J., Jeffers, S. V.,
Paletou, F. 2007b, MNRAS, 380, 1297

\bibitem{} Donati, J.-F., Jardine, M. M., Gregory, S. G., Petit, P., Paletou, F.,
Bouvier, J., Dougados, C., M\`{e}nard, F., Cameron, A. C., Harries, T. J.,
Hussain, G. A. J., Unruh, Y., Morin, J., Marsden, S. C., Manset, N., Auri\`{e}re, M., Catala, C., Alecian, E. 2008, MNRAS, 386, 1234

\bibitem{} Goodson, A.P., \& Winglee, R. M., 1999, ApJ, 524, 159

\bibitem{goo97} Goodson, A.P., Winglee, R. M., \& B\"ohm, K.-H. 1997, ApJ,
489, 199

\bibitem{goo99} Goodson, A.P., B\"ohm, K.-H., Winglee, R. M. 1999, ApJ, 524,
142

\bibitem{} Jardine, M., Collier Cameron, A., \& Donati, J.-F. 2002,
MNRAS, 333, 339

\bibitem{} Korneich, D.A., \& Lovelace, R.V.E. 2008, ApJ, 681, 104

\bibitem{} Long, M., Romanova, M. M., \& Lovelace, R. V. E. 2007, MNRAS, 374, 436

\bibitem{} --------- 2008, MNRAS, 386, 1274

\bibitem{} Long, M., Romanova, M.M., Lamb, F.K., 2009, MNRAS, arXiv:0911.5455

\bibitem{} Long, M., Romanova, M. M., Lamb, F., Kulkarni, A.K., \& Donati, J.-F. 2010, MNRAS, in preparation

\bibitem{} Lovelace, R.V.E., Berk, H.L., \& Contopoulos, J. 1991, ApJ, 379,
696

\bibitem{} Lovelace, R.V.E., Romanova, M.M., \& Bisnovatyi-Kogan, G.S.
1999, ApJ, 514, 368

\bibitem{} Lovelace, R.V.E., Rothstein, D.M., \& Bisnovatyi-Kogan, G.S.
2009, ApJ, 701, 885

\bibitem{} Pariev, V.I., \& Colgate, S.A. 2007, ApJ, 658, 129

\bibitem{} Pariiev, V.I., Colgate, S.A., \& Finn, J.M. 2007, ApJ, 658, 129

\bibitem{} Romanova, M.M., Ustyugova, G.V., Koldoba, A.V., \&
Lovelace, R.V.E. 2005, ApJ, 635, 165L

\bibitem{} Romanova, M.M., Ustyugova, G.V., Koldoba, A.V., \&
Lovelace, R.V.E. 2009, MNRAS, 399, 1802 (R09)

\bibitem{} Rothstein, D.M., \& Lovelace, R.V.E. 2008, ApJ,
677, 1221


\bibitem{} Shakura, N.I., \& Sunyaev, R.A. 1973, A\&A, 24, 337

\bibitem{} Shu, F., Najita, J., Ostriker, E., Wilkin, F., Ruden, S.,
Lizano, S. 1994, ApJ, 429, 781

\bibitem{} Smirnov, D. A., Lamzin, S. A., Fabrika, S. N., \& Chuntonov, G. A., 2004, Astron. Lett., 30, 456

\bibitem{} Tsygan, A.I. 2007, Astron. Rep., 51, 97

\bibitem{} Ustyugova, G.V., Koldoba, A.V., Romanova, M.M., \& Lovelace, R.V.E. 2006, ApJ, 646, 304 (U06)

\bibitem{} von Rekowski B., \& Brandenburg A., 2006, Astron. Nachr., 327, 53

\bibitem{} Wang, J.C.L., Sulkanen, M.E., \& Lovelace, R.V.E. 1992,
ApJ, 390,  46


\bibitem{} Woitas, J., Ray, T.P., Bacciotti, F., Davis, C.J., \& Eisl\"offel, J.
2002, ApJ, 580, 336



\end{thebibliography}
